\documentclass[preprint]{vgtc}               % preprint
\graphicspath{{figures/}{pictures/}{images/}{./}} % where to search for the images

\usepackage{times}                     % we use Times as the main font
\usepackage{tabu}                      % only used for the table example
\usepackage{booktabs}                  % only used for the table example
\usepackage{lipsum}                    % used to generate placeholder text
\usepackage{mwe}                       % used to generate placeholder figures

\usepackage{mathptmx}                  % use matching math font
\usepackage{url}
\usepackage{enumitem}

\onlineid{1150}

\vgtccategory{Research}

\vgtcinsertpkg

\usepackage{xurl}
\usepackage{xparse}

\NewDocumentCommand{\codeword}{v}{%
\texttt{\textcolor[HTML]{386594}{#1}}%
}

\newcommand{\tool}{PyGWalker~}
\newcommand{\toole}{PyGWalker}
\newcommand{\gw}{Graphic Walker~}
\newcommand{\gwe}{Graphic Walker}
\title{\toole: On-the-fly Assistant for Exploratory Visual Data Analysis}

\usepackage{scalerel}
\usepackage{tikz}
\usetikzlibrary{svg.path}

\definecolor{orcidlogocol}{HTML}{A6CE39}
\tikzset{
  orcidlogo/.pic={
    \fill[orcidlogocol] svg{M256,128c0,70.7-57.3,128-128,128C57.3,256,0,198.7,0,128C0,57.3,57.3,0,128,0C198.7,0,256,57.3,256,128z};
    \fill[white] svg{M86.3,186.2H70.9V79.1h15.4v48.4V186.2z}
                 svg{M108.9,79.1h41.6c39.6,0,57,28.3,57,53.6c0,27.5-21.5,53.6-56.8,53.6h-41.8V79.1z M124.3,172.4h24.5c34.9,0,42.9-26.5,42.9-39.7c0-21.5-13.7-39.7-43.7-39.7h-23.7V172.4z}
                 svg{M88.7,56.8c0,5.5-4.5,10.1-10.1,10.1c-5.6,0-10.1-4.6-10.1-10.1c0-5.6,4.5-10.1,10.1-10.1C84.2,46.7,88.7,51.3,88.7,56.8z};
  }
}

\newcommand\orcidicon[1]{\href{https://orcid.org/#1}{\mbox{\scalerel*{
\begin{tikzpicture}[yscale=-1,transform shape]
\pic{orcidlogo};
\end{tikzpicture}
}{|}}}}

\usepackage{hyperref}

\author{Yue Yu~\orcidicon{0000-0002-9302-0793}\thanks{e-mail: yue.yu@connect.ust.hk. This work was done during Yue Yu's internship at Kanaries Data.}\\
\parbox{1.2in}{\scriptsize \centering Hong Kong University of Science and Technology \\ Kanaries Data} %
\and Leixian Shen~\orcidicon{0000-0003-1084-4912}\thanks{e-mail: lshenaj@connect.ust.hk}\\ %
     \scriptsize \parbox{1.2in}{Hong Kong University of Science and Technology} %
\and Fei Long~\orcidicon{0009-0006-0090-6197}\thanks{e-mail: feilong@kanaries.net}\\ %
     \scriptsize Kanaries Data %
\and Huamin Qu~\orcidicon{0000-0002-3344-9694}\thanks{e-mail: huamin@cse.ust.hk}\\ %
     \scriptsize \parbox{1.2in}{Hong Kong University of Science and Technology} %
\and Hao Chen~\orcidicon{0009-0002-6440-5296}\thanks{e-mail: haochen@kanaries.net. Hao Chen is the corresponding author.}\\
     \scriptsize Kanaries Data %
    }

\teaser{
  \centering
  \includegraphics[width=\linewidth]{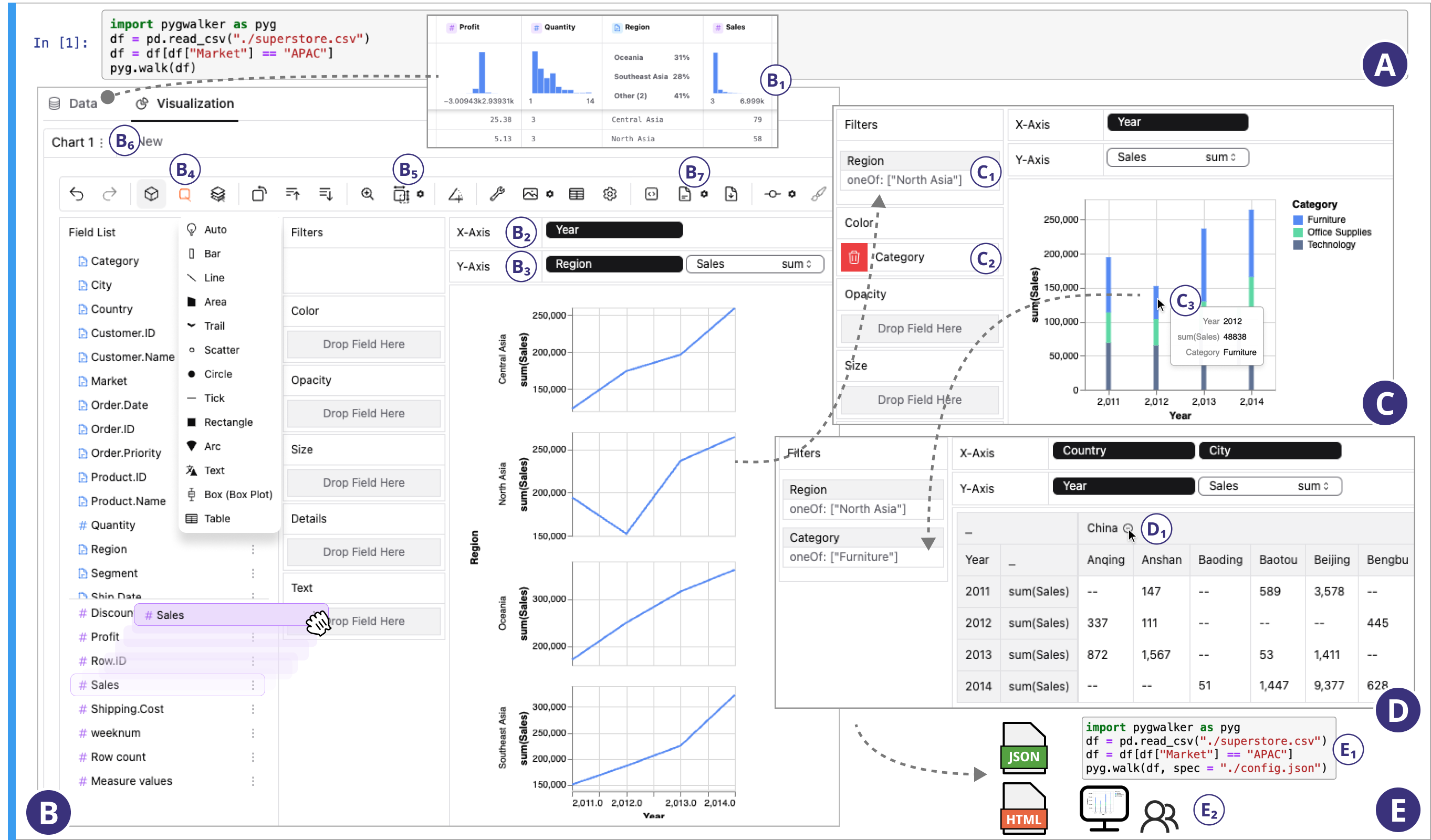}
  \caption{An example walkthrough with \tool in a Jupyter Notebook. The analyst initially loads the dataset into \tool to activate the interface with one line of code ($A$). 
  The dataset overview is provided under the Data tab ($B_1$).
  Under the Visualization tab, the analyst creates multi-faceted charts ($B$) by dragging data columns to the X-Axis ($B_2$) and Y-Axis ($B_3$) and selecting the mark type ($B_4$). 
  Filters are further added to narrow the dataset ($C_1$), and the color encoding ($C_2$) can be diversified for varied visualizations ($C$). 
  Details can be viewed by hovering over chart elements ($C_3$).
  In addition to charts, pivot tables can be constructed ($D$) with collapsible hierarchical headers ($D_1$). 
  Finally, the visualizations generated in the exploratory analysis process can be documented ($E$).
  The analyst adjusted and chart style ($B_5$), renamed charts ($B_6$), and exported them ($B_7$) as a JSON specification ($E_1$) and an HTML file of the notebook ($E_2$), which can be shared and hosted to reproduce the analysis process ($E_2$), accessible at this \href{https://kanaries.net/gallery/pygwalker/notebooks/superstore.html}{link}.
  }
  \label{fig:teaser}
}

\abstract{
Exploratory visual data analysis tools empower data analysts to efficiently and intuitively explore data insights throughout the entire analysis cycle. However, the gap between common programmatic analysis (e.g., within computational notebooks) and exploratory visual analysis leads to a disjointed and inefficient data analysis experience. To bridge this gap, we developed \toole, a Python library that offers on-the-fly assistance for exploratory visual data analysis. It features a lightweight and intuitive GUI with a shelf builder modality. Its loosely coupled architecture supports multiple computational environments to accommodate varying data sizes. Since its release in February 2023, \tool has gained much attention, with 612k downloads on PyPI and over 10.5k stars on GitHub as of June 2024. This demonstrates its value to the data science and visualization community, with researchers and developers integrating it into their own applications and studies. 
} % end of abstract


\begin{document}


\firstsection{Introduction}

\maketitle

Exploratory visual data analysis tools can help data analysts effectively and intuitively explore data insights and guide further analyses. This process is iterative and spans the entire data analysis cycle.
However, when analysts engage in programming, such as within computational notebooks, numerous data variables are generated. 
Mapping these variables into visual representations that align with the user's intentions and supporting exploratory visual analysis is non-trivial, especially for those who are unfamiliar with visual design. Analysts are often faced with the choice of either exporting the data and importing it into existing exploratory visual data analysis tools (e.g., Voyager~\cite{Wongsuphasawat2016c} and DeeyEye~\cite{Luo2018}) or relying on complex programming to repeatedly visualize their data. The gap between programmatic analysis and exploratory visual analysis leads to a disjointed data analysis experience.

To address these issues, we developed \toole, a Python library that offers on-the-fly assistant for exploratory visual data analysis in computational notebooks,
% designed to seamlessly integrate exploratory visual data analysis into data analysis workflows, 
available as open source at \href{https://github.com/Kanaries/pygwalker}{github.com/Kanaries/pygwalker}. 
\tool enables users to effortlessly invoke a lightweight exploratory visual data analysis tool with just a single line of code, as shown in \cref{fig:teaser}. The intuitive graphic user interface (GUI) features a shelf builder modality~\cite{Satyanarayan2019a}, allowing users to intuitively drag and drop variables onto visual channels and dynamically experiment with various visual representations. Furthermore, its architecture, which decouples interaction, computation, and rendering, supports the integration of multiple computational environments, such as JavaScript, Python kernel, and external databases, to accommodate varying data sizes.

\toole's rapid adoption underscores its value to the data science and visualization community. 
% Since its release in February 2023, it has garnered significant attention, with a total of 612k downloads on PyPI~\cite{pypi} and over 10.5k stars on GitHub as of June 2024. 
Community members have enthusiastically contributed tutorials and demos~\cite{youtube1, youtube2, medium3, medium1, youtube3, medium2}, with researchers and developers integrating \tool into their own applications and studies.

\section{Related Work}
\tool draws upon existing works about exploratory visual data analysis and assistance in computational notebooks.

\textbf{Exploratory Visual Data Analysis:}
Data analysis often involves users continuously exploring insights through iterations and drilling down. Existing exploratory visual data analysis tools can help users effectively and intuitively complete these tasks. For example, Voyager~\cite{Wongsuphasawat2016c} uses statistical and perceptual measures to recommend suitable charts and supports faceted browsing. MEDLEY~\cite{Pandey2022} and TaskVis~\cite{Shen2021,shendata} support user-intention-driven visual data analysis. DMiner~\cite{Lin2023} and MultiVision~\cite{Wu2021e} automatically generate multi-view dashboards based on user data. However, these are all independent analysis systems. The separation between coding and exploratory visual data analysis requires users to switch between two modes, hindering a seamless data analysis process.

\textbf{Assistance in Computational Notebooks:}
Computational notebooks are a popular data analysis environment, and many studies have developed corresponding assistive features for visual analysis and storytelling. For example, Lux~\cite{Lee2021e} is an always-on framework that helps users quickly preview insights in their data. Notable~\cite{Li2023} can automatically convert data facts that users are interested in during their analysis process into slides to facilitate data communication. InkSight~\cite{Lin2023} allows users to sketch interesting data insights directly on visualizations and then uses LLMs to generate descriptions. B2~\cite{Wu2020a} uses data queries as a shared representation to bridge the gap between code and interactive visualizations, automatically generating dashboards. Although these tools can quickly help users generate data analysis and communication results, they all lack support for on-the-fly process-oriented exploratory visual analysis.
% \textbf{Visualization Recommendation}

\section{\tool}
In this section, we will first give an overview of \toole, and then walk through a usage scenario to demonstrate the interaction workflow for users. Next, we will discuss the computation engine behind \tool and introduce the rendering of visualizations.

\begin{figure}[t]
    \centering
    \includegraphics[width=\linewidth]{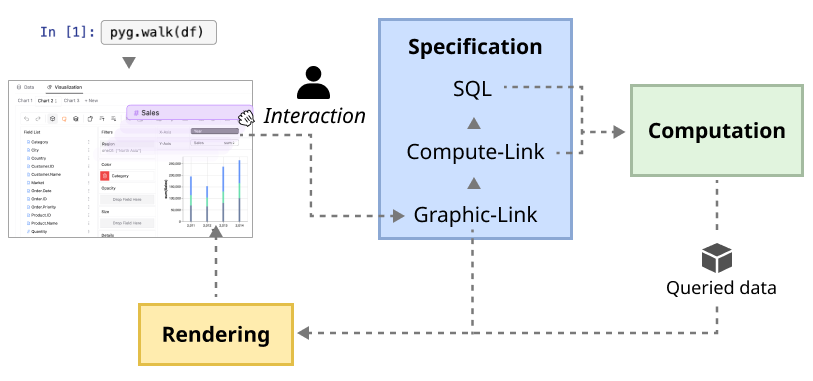}
    \vspace{-0.7cm}
    \caption{Overview of \toole.}
    \label{fig:overview}
    \vspace{-0.5cm}
\end{figure}

\subsection{Overview}
The main workflow is as follows: users can easily insert a line of code to call \tool during data analysis, which will activate the interactive GUI (named \gwe) and automatically load the relevant data. To represent the user's intent information and data information during interactions on the GUI, we have designed a declarative scheme, called Graphic-Link (\cref{sec:GWlink}). Furthermore, to support various computing paradigms and environments (e.g., browsers, databases, cloud computing, etc.), we have also derived a computation-driven scheme, called Compute-Link (\cref{sec:computelink}), from Graphic-Link, aiming to decouple view data processing from visualization specifications. Compute-Link allows the generating of user-intent-based view data or SQL queries. The aggregated data from these queries, as well as Graphic-Link, is then rendered into interactive visualizations in real time. As the user iteratively conducts exploratory visual analysis on \gwe, the corresponding specification is continuously updated.

% \subsection{Interaction Design}

\subsection{Usage Scenario}
% To motivate the interaction design of \toole, we use a public dataset on Kaggle\footnote{\url{https://www.kaggle.com/datasets/apoorvaappz/global-super-store-dataset}} to illustrate how a sales analyst in a superstore enterprise, Alice, can use \tool to seamlessly integrate visual exploratory data analysis into her workflow in the programming environment, as shown in ~\cref{fig:teaser}.

Imagine Alice, a data analyst in a superstore enterprise, is assigned a task to analyze the global superstore dataset~\cite{kaggle}. \cref{fig:teaser} illustrates how she integrates \tool into her programming workflow to seamlessly conduct exploratory visual data analysis.

In her Jupyter Notebook, Alice begins by importing the global sales dataset from a CSV file, applying a filter to narrow her analysis to the Asia Pacific market. 
She then initiates \tool by executing \codeword{pyg.walk(df)} within her notebook ($A$), and \toole's interactive GUI directly appears in the output cell.
After glancing over the dataset overview under the Data tab ($B_1$), she turns to the Visualization tab and starts her exploration.
Alice is interested in the sales over years of different regions for a broad picture. 
Therefore, she drags the \textit{Year} field from the Field List and drops to the X-Axis shelf ($B_2$), and \textit{Region} and Sum of \textit{Sales} fields to the Y-Axis shelf ($B_3$), using the \textit{Line} mark type ($B_4$).
The interface responds by displaying four line charts ($B$), faceted by region, showing a significant drop in sales for North Asia in 2012, prompting Alice to investigate this anomaly more deeply.

Alice thus removes the \textit{Region} from the X-Axis and adds a filter by dragging the \textit{Region} field into the Filters panel, filtering for ``North Asia'' ($C_1$).
To explore sales by category, she drags the \textit{Category} field to the Color encoding shelf ($C_2$) and switches the mark type to \textit{Bar}.
The resulting bar chart ($C$) shows a notably sharp decline in the blue segment in 2012. Hovering on the segment ($C_3$), the tooltip shows the segment is the ``Furniture'' category.

Curious about the specific cities contributing to the decline, Alice shifts to a more granular-level analysis by examining the exact numbers of furniture sales across different cities over the years.
She applies a filter for ``Furniture'', sets the mark type to Table, and organizes the data hierarchically with \textit{Year} on the Y-Axis and \textit{Country} and \textit{City} on the X-Axis.
This creates a pivot table that groups cities within their countries by year ($D$).
As there are hundreds of cities, she first collapses the country-level headers ($D_1$) and notes that between 2011 and 2012, sales in China and South Korea dropped significantly. By expanding the headers of these countries, she identifies major cities like Beijing, Jining, and Seoul, which showed drastic sales declines from robust figures in 2011 to near-zero in 2012, pinpointing them as the primary sources of the downturn.

Alice finally documents and shares her findings ($E$). 
After adjusting the chart style ($B_5$) and renaming the chart tabs ($B_6$), she exports the visualization specifications into a JSON file \codeword{config.json} ($B_7$) and shares it with her colleagues responsible for the supply chain in the affected cities. 
They can replicate her visualizations by loading the specification using \codeword{pyg.walk(df, spec="./config.json")} in their own notebooks ($E_1$). 
Additionally, Alice exports the entire notebook as an HTML file\footnote{\url{https://kanaries.net/gallery/pygwalker/notebooks/superstore.html}} and hosts it online, enabling colleagues without a programming environment to view her exploration steps in different tabs and interact with her visualizations ($E_2$). 
% The generated interactive visualizations showing her exploration steps in different tabs.

\subsection{Information Representation}\label{sec:GWlink}
% \subsubsection{Graphic-Link}
% We devise the Graphic-Link inspired by the Grammar of Graphics~\cite{Wilkinso} as the visualization specification.
After the user loads their data with \tool and interacts on the \gw GUI, the data and user's analysis intents through interactions are stored in a declarative specification, named Graphic-Link\footnote{\url{https://graphic-walker.kanaries.net/chartinfo.json}}.
It mainly encodes the following primitive types:
% Graphic-Link systematically captures the user's analysis intent and relevant data characteristics during interactions within the \gw GUI, with three key components:

% \begin{itemize}[itemsep=0.05pt, leftmargin=*]
    % \item 
\textbf{Data Characteristic:}
% provide a principled way to map data columns to visual channels. 
Graphic-Link encodes data columns and their automatically inferred semantic information.
Each data column is assigned a semantic data type (nominal, ordinal, or quantitative) and an analytic type, which is either a ``dimension'' (categorical descriptors that segment data) or a ``measure'' (quantifiable metrics)~\cite{Janus2009}. This helps that visualizations faithfully depict the inherent relationships present in the data.

% Graphic-Link encodes the data columns and their automatically inferred semantic information. Each data column has a semantic data type (nominal, ordinal, or quantitative) and an analytic type, ensuring visualizations stay true to the underlying data relationships. Each data column has an analytic type, either as ``dimension'' (categorical or temporal data for grouping) or ``measure'' (quantitative fields requiring aggregation). Further, it supports field transformations  (e.g., logarithms, binning) and sorting options. 
    
% \item 
\textbf{Data Transformation:} 
Graphic-Link keeps a record of users' data transformation operations. Users can interactively define filtering criteria, enabling the specification of value sets or ranges for data columns. Moreover, users can explicitly specify aggregation functions such as ``sum'' and ``average'', and performing field transformations (such as logarithms or binning) will result in a new transformed data column. Users can also sort and stack their data with different styles.
% to guide the generation of Compute-Link (refer to Section~\ref{sec:computelink}). 
% Furthermore, field transformations (e.g., logarithms, binning) and sorting options, which contribute to an enriched experience of data exploration and analysis.

% Users can define filtering criteria based on data columns, specifying value sets or ranges (for quantitative fields). They can also explicitly define aggregation functions (e.g., ``sum'' and ``average'') applied to measures to inform the generation of Compute-Link (Section~\ref{sec:computelink}). Moreover, Graphic-Link offers additional functionalities such as field transformations (e.g., logarithms, binning) and sorting options to enhance data exploration and analysis.

% \item 
\textbf{Visual Encoding:} 
Graphic-Link captures and records users' interactions with the GUI, which represents their intentions for visual data analysis.
Users can define the fundamental geometric primitives employed in the visualization, such as ``line,'' ``bar,'' ``point,'' and more. Additionally, to enhance the exploration process, a default mark type is automatically generated based on heuristic rules inspired by existing automatic visualization creation systems~\cite{showme, Wongsuphasawat2016c}.
Furthermore, users can intuitively assign their data columns to various visual channels, including the X-axis, Y-axis, color, size, shape, opacity, and more, by simply dragging and dropping them.
% Furthermore, it supports automatic faceting when multiple dimensions are organized hierarchically along an axis for effective comparisons, and users can specify the independence of axes when multi-facet is effective.
Furthermore, it supports automatic faceting when multiple dimensions are hierarchically organized along an axis. 
% This enables users to discern patterns and relationships across different facets of the data.
% These mappings, along with encodings, determine the visual representation of data elements.
    
% \item 
\textbf{Style and Configuration:} To facilitate cross-platform viewing and sharing, Graphic-Link also encodes additional configurations related to coordinate systems (chart or geographic), layout modes, chart styles, scale ranges, color palettes, etc. This ensures that visualizations remain consistent and accurate across different platforms and devices, preserving the intended visual aesthetics.
% interactive settings and formatting, so that users can create more tailored and informative visualizations by configurations such as the chart size, color palette, and scale range for the color or opacity encoding, etc.
% \end{itemize}

Graphic-Link provides a comprehensive representation of data characteristics and user interaction information, establishing a foundation for subsequent computation and rendering processes.

\subsection{View Data Computation}\label{sec:computelink}
To enable adaptability across diverse computational paradigms, we additionally introduce Compute-Link\footnote{\url{https://github.com/Kanaries/graphic-walker/blob/main/computation.md}}, a computation-oriented specification that separates view data processing from visualization specifications. While Graphic-Link aims to declare the visualization contents, Compute-Link extracts the necessary view data for a given visualization from the comprehensive Graphic-Link specification.  
The derivation from Graphic-Link to Compute-Link mirrors the user's process of mapping data columns to visual channels in the GUI. 
This transformation is guided by algebraic rules in the Grammar of Graphics~\cite{Wilkinson} (i.e., cross, nest, and blend operators), which determines how different data attributes are combined and represented to compute the view data, particularly for constructing pivot tables and multi-faceted charts.

Compute-Link consists of a series of data queries that describe how to compute the view data from the raw data:
% Compute-Link encompasses a set of structured query steps tailored to meet the specific demands of user-defined computational tasks:
\textit{Filter Query} selects specific data subsets based on users' defined criteria;
\textit{Transform Query} applies calculations and transformations to the data columns, such as calculating the logarithm or binning;
\textit{View Query} structures the data into the desired format for visualization, which can be aggregated by different operations or directly structured as raw;
and \textit{Sort Query} orders the result data.

% Compute-Link is a computation-oriented specification, decoupling data processing from visualization specification. While Graphic-Link is designed to declare the contents of the visualization, Compute-Link derives the required view data for a given visualization from the complete specification (Graphic-Link) which describes what is depicted on the visualization, thus allowing flexibility across different computational paradigms. Compute-Link includes a series of structured query steps adapting to the specific requirements of the user-defined analytical task:
% The derive from Graphic-Link to Compute-Link involves the generation of a computation schema based on user interactions.

% \begin{itemize}[itemsep=0.05pt, leftmargin=*]
%     \item \textbf{Filter} selects specific data subsets based on defined criteria.
%     \item \textbf{Transform} applies calculations and transformations to the data columns, such as calculating the logarithm or binning.
%     \item \textbf{View} structures the data into the desired format for visualization, which can be aggregated by different operations or directly structured as raw.
%     \item \textbf{Sort} orders the data based on user-defined criteria.
% \end{itemize}

\begin{figure}[t]
    \centering
    \includegraphics[width=\linewidth]{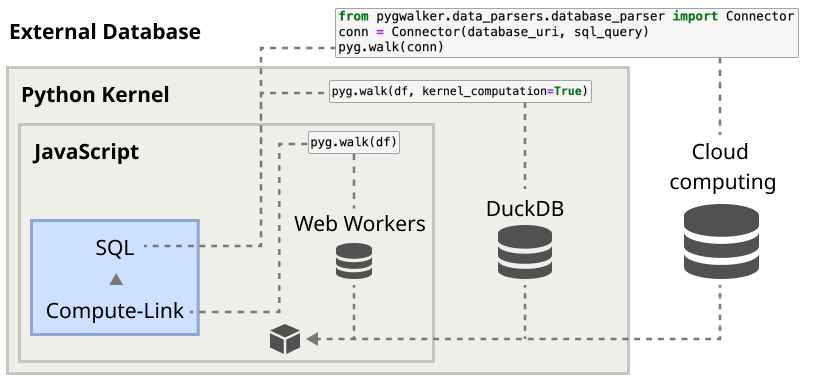}
    \vspace{-0.7cm}
    \caption{Computation mechanisms of \toole.}
    \label{fig:computation}
    \vspace{-0.5cm}
\end{figure}

This separation not only maintains a clear decouple from the visualization specification but flexibly supports a wide array of computational modes and environments such as web browsers, databases, and cloud computing platforms, as illustrated in Fig.~\ref{fig:computation}. The following illustrates some computational scenarios:

\textbf{JavaScript Computation:}
By default, when a user executes \codeword{pyg.walk(df)}, the computation occurs within the JavaScript environment of the application, utilizing a pipeline of web workers for asynchronous data processing. 
These workers parse and execute the series of operations defined in Compute-Link, ultimately producing the finalized view data.

\textbf{Python Kernel Computation:}
While JavaScript computation may be limited to small-scale computation, users can opt for Python kernel computation by setting \codeword{kernel_computation=True} for more substantial datasets that exceed the capabilities of JavaScript.
This mode parses Compute-Link specifications into analytical SQL queries to be processed by the DuckDB engine~\cite{duckdb}, an open-source analytic database with high computation performance, in the Python Kernel.
The results are then conveyed back to the JavaScript environment for rendering.

\textbf{External Database Computation:}
For users who need to analyze large datasets stored in non-local databases or online data warehouses, \tool also offers an external computation capability.
To utilize this functionality, users can import \codeword{Connector} class from the \codeword{database_parser} module and create a Connector object (e.g., \codeword{conn}) by supplying the SQLAlchemy connection URI and a SQL query to retrieve the dataset.
Once configured, the analysis can be conducted directly on the external database by executing \codeword{pyg.walk(conn)}.
Similar to kernel computation, analytical SQL queries will be generated and sent to the external database, and the results are then fetched and passed to the JavaScript environment.

% 基于《The Grammar of Graphics》(Leland Wilkinson 著)一书所提的图形理论。该理论是一套用来描述所有统计图形深层特性的语法规则，该语法回答了『什么是统计图形』这一问题，以自底向上的方式组织最基本的元素形成更高级的元素。

\subsection{Visualization Rendering}
Graphic-Link specifications, combined with the computed view data, are finally translated into visual representations, which support the rendering of different visualization types:
\textit{General Charts:} For common chart types (e.g., lines, bars, points, etc.), Graphic-Link is initially converted into a Vega-Lite specification~\cite{vegalite}, leveraging the Vega-Lite rendering engine for its expressiveness, widespread usage, and adherence to the principles of the Grammar of Graphics~\cite{Wilkinson}.
\textit{Pivot Tables:} When a table mark type is chosen, a pivot table is directly rendered to effectively display aggregated data.
\textit{Geographic Visualizations:} In the case of drawing a geographic visualization, an OpenStreetMap base layer will be established, overlaid by encodings that are mapped to longitude and latitude, ensuring accurate spatial representation of data elements.

\subsection{Cross-Platform}
% \tool extends its reach beyond Jupyter Notebooks, supporting data analysts across a variety of platforms and languages.
In addition to providing support for Jupyter Notebooks, \tool expands its capabilities to cater to data analysts working on different platforms and utilizing various programming languages.

% \textbf{Notebook Environments and Python Applications:} 
\tool seamlessly integrates with popular notebook-based analysis environments like Colab and Kaggle Notebook, promoting collaboration and knowledge sharing.
% Furthermore, with the ability to be embedded into Python applications supporting HTML rendering (such as ), 
Furthermore, \tool can also be embedded into Python frameworks that support HTML rendering, such as Streamlit~\cite{streamlit} and Gradio~\cite{gradio}. 
This empowers developers to create bespoke visualization analysis interfaces within their own applications swiftly.
% , enabling the effortless addition of interactive data exploration capabilities to their applications.
% \tool facilitates the rapid creation of custom visualization analysis interfaces. This allows developers to easily add interactive data exploration capabilities to their applications.

% \textbf{Integration in Other Environments:} 
Owing to the flexible design of the architecture of \toole, the interactive GUI, \gwe\footnote{\url{https://github.com/Kanaries/graphic-walker}}, can be viewed as a standalone TypeScript-based React library to enable flexible integration into other environments, such as web applications and other programming languages.
For example, recognizing the diverse data science landscape, we've developed a similar package for the R programming environment\footnote{\url{https://github.com/Kanaries/GWalkR}}. This opens the door to exploratory visual analysis for a large community of R-focused data scientists.

\section{Release and Usage}
\tool\footnote{\url{https://github.com/Kanaries/pygwalker}} was first released and open-sourced in February 2023. As of June 2024, it has gained significant popularity, with over 612k total downloads, approximately 90k monthly downloads, and over 10.5k stars on GitHub. These quantitative metrics indicate that \tool has established a strong presence in the developer community.
To delve deeper into \toole's real-world impact, we systematically analyzed how developers and researchers utilize the tool. Our findings are based on two key sources:
(1) Metadata from 536 public GitHub repositories listing \tool as a dependency and (2) 7 research articles on Google Scholar containing the keyword ``\toole'' in their works.

% \begin{figure}[t]
%     \centering
%     \includegraphics[width=\linewidth]{figures/bi_cat.png}
%     \caption{Interface of BICat~\cite{github_bicat}, an open-source project on GitHub that features \tool in a Streamlit-based web application.}
%     \label{fig:bi_cat}
%     \vspace{-8px}
% \end{figure}

% \textbf{Empowering Python Applications with Visualization:} 
The analysis of GitHub repositories revealed some interesting insights. For example, among 536 repositories using \toole, 358 (66.8\%) are Python applications, 138 (25.7\%) are Jupyter Notebooks, and the remaining 40 (7.5\%) utilize other languages like HTML and JavaScript.
We also found 101 repositories (18.8\%) explicitly mentioning frameworks like ``Streamlit,'' ``Django,'' and ``Flask''.
This suggests \toole's adaptability for seamless integration into web applications, extending its value to the open-source developer community beyond traditional data analysis workflows.
Looking into some detailed cases, we found developers typically leverage \tool as an out-of-the-box exploratory dashboard to support dynamic data visualization.
For example, BICat~\cite{github_bicat} is a Streamlit-based application that integrates \tool and ChatGPT~\cite{chatgpt} to enable users to interactively explore an uploaded dataset by natural language.
% , as shown in Fig.~\ref{fig:bi_cat}. 
Similarly, Diagnostic Expert Advisor~\cite{polzin2023diagnostic} juxtaposes \tool with predefined plots, offering users both guided views and the freedom of self-directed exploration.
These findings underscore how \toole's convenient and seamless integration empowers developers to build web applications moving beyond static visualizations.

% \textbf{Supporting Research Across Domains:} 
While we understand GitHub may not fully reflect the extent of \toole's use in notebooks due to data privacy considerations, we found compelling evidence of its adoption in research.
Researchers from diverse fields, including meteorology~\cite{eltaweel2024prediction}, traffic~\cite{lo2023optimisation}, biomechanics~\cite{mokhtarzadeh2023streamlining}, and waste management~\cite{github_waste}, have embraced \toole, showcasing its potential effectiveness across disciplines.
Besides research applications, \tool is also introduced in data science education programs. 
For instance, the book chapter \textit{Data Visualization for Business Intelligence}~\cite{sukhdeve2023data} features \tool for data visualization in business intelligence, stating it \textit{``simplifies the data analysis and data visualization workflow.''}
Furthermore, Python crash courses like better-py/learn-py~\cite{github_learn} integrate \toole, emphasizing its user-friendly nature and low-code approach.
These cases show \toole's capabilities to enable learners to explore datasets quickly and gain visual insights, enhancing their data analysis skills.
% \textbf{Possibilities in Data Science Education}:

\section{Discussion}
We discuss the limitations of \tool and potential future directions to enhance it.

\textbf{Expand Visualization Capabilities:}
\toole's architecture is intentionally designed to separate visualization rendering from information representation and data processing, allowing flexible integration of diverse rendering libraries and engines.
Currently, \tool primarily relies on Vega-Lite, benefiting from its expressiveness and ease of use. 
However, we acknowledge that for specialized or interactive visualizations with complex requirements, a lower-level library like D3~\cite{d3} may be indispensable.
In the future, we plan to empower users with greater customization and control, allowing them to create a wider range of visual narratives with their data, including support for more visualization types~\cite{Shen2022} and narrative types~\cite{Idb}.

\textbf{Design Intelligent Interactions:}
While our current interaction design is intuitive, it does require users to have some understanding of the tool-specific operations to produce effective visualizations.
% However, integrating large language models (LLMs) offers a more natural and intelligent interaction mode, 
One promising direction is integrating natural language interfaces with the help of large language models (LLMs), allowing users to freely talk their analysis intents to the system and directly generate corresponding results~\cite{Shen2021a}. 
Furthermore, \toole's interactive interface provides opportunities for users to explore and refine the visualizations generated by LLMs~\cite{vistalk}, creating a feedback loop. 
To leverage these capabilities, we are actively developing a feature that translates natural language descriptions and dataset metadata into Graphic-Link specifications.

\textbf{Leverage \tool as a Learning Tool:}
Previous research underscores the benefits of interactive notebooks for teaching data visualization over traditional slides~\cite{lo2019}, shedding light on the potential for \tool as an effective education tool.
Currently, \tool only focuses on exporting the final visualization output for further use.
In the future, we plan to enhance the analytic provenance that captures the visualization creation process~\cite{north2011}.
This would allow educators to share step-by-step walkthroughs of their visual data analysis workflows, enabling students to learn by actively retracing the expert's reasoning and experimentation.

\section{Conclusion}
This paper proposes \toole, which bridges the gap between programmatic analysis and exploratory visual analysis tools. Its user-friendly GUI and loosely coupled architecture have gained considerable attention from the data science and visualization community, with users actively contributing tutorials and developers integrating it into their own applications. We will continually enhance and expand the tool's functionality, contributing to the open-source community. We also encourage more users to join in its development and inspire further interesting and practical research.

% \newpage
% \noindent
% \textbf{Supplemental Material:}
% To provide deeper insights into the community engagement and adoption of \toole, we include the following CSV files in the PCS system:
% \begin{itemize}
%     \item Community-Created Videos (21 entries): This file showcases the videos that introduce \tool on YouTube, including video titles, creators, dates, languages, and links.
%     \item Dependent Repositories Metadata (536 entries):  This file provides information on GitHub repositories using \toole, including repository names, links, descriptions, programming languages, topics, and star counts.
% \end{itemize}

\acknowledgments{
The authors wish to thank all of the contributors from our open-source community who support \tool and related projects.}

\bibliographystyle{abbrv-doi}

\bibliography{main}

\end{document}